\documentclass[aps,prb,amsmath,amssymb,twocolumn,superscriptaddress,showpacs,floatfix]{revtex4-1}

\usepackage{graphicx}
\usepackage{dcolumn}
\usepackage{bm}

\usepackage{amsmath}
\usepackage{amssymb}

\DeclareMathOperator{\eps0}{\varepsilon_0}
\DeclareMathOperator{\epsi}{\varepsilon}

\begin{document}

\title{Influence of image forces on the electron transport in ferroelectric tunnel junctions}

\author{O.~G.~Udalov}
\affiliation{Department of Physics and Astronomy, California State University Northridge, Northridge, CA 91330, USA}
\affiliation{Institute for Physics of Microstructures, Russian Academy of Science, Nizhny Novgorod, 603950, Russia}
\author{I.~S.~Beloborodov}
\affiliation{Department of Physics and Astronomy, California State University Northridge, Northridge, CA 91330, USA}

\date{\today}


\begin{abstract}
We study influence of image forces on conductance of ferroelectric tunnel junctions.
We show that the influence of image forces is twofold: i) they enhance the electro-resistance 
effect due to polarization hysteresis in symmetric tunnel junctions at non-zero bias and 
ii) they produce the electro-resistance effect due to hysteresis of dielectric permittivity 
of ferroelectric barrier. We study dependence of ferroelectric tunnel junction conductance
on temperature and show that image forces lead to strong conductance variation with temperature.
\end{abstract}

\maketitle
\section{Introduction}

Recent progress in fabrication techniques allows creating nanometer scale ferroelectric (FE) films~[\onlinecite{Thompson2004,Ghosez2005,Xi2006,Zlatkin1998}] and
FE tunnel junctions (FTJ) where metallic leads are separated by tunnel barrier made of FE material [\onlinecite{Chu2016, Barthelemy2010,Katiyar2014,Okamura2014,Alexe2012,Zhong2016}]. The most promising and important property of FTJ is the electro-resistance (ER) effect meaning the dependence of the FTJ linear conductance on the polarization direction of the FE layer.
The ER effect can be used for non-volatile memory applications~[\onlinecite{Bibes2014,Gruverman2012,Grollier2014,Chappert2014}]. Several phenomena were considered to be responsible for ER effect in FTJ such as barrier thickness variation due to strain in FE~[\onlinecite{Woo2009,Waser2005}], variation of band structure of FE layer~[\onlinecite{Waser2005}] and the appearance of surface charges at the FE/metal interfaces~[\onlinecite{Tsymbal2005,Kohlstedt2006}]. It was demonstrated that the last
mechanism is the strongest one leading to the giant ER (GER) effect up to 1000\%.
ER effect linear in FE polarization appears only in 
asymmetric FTJs with essentially different metallic leads or in metal/FE/semiconductor
structures~[\onlinecite{Ming2013, Gruverman2013}]. 

In symmetric TJ with the leads made of the same metal the ER effect appears
for non-linear conductance at finite bias.
In this case the contribution to the conductance contains a combination of
polarization and the bias voltage. The magnitude of the effect is not as high
as GER in asymmetric FTJ, but may reach several tens of percent~[\onlinecite{Waser2005}] 
which is comparable to magnetoresistance effect in magnetic tunnel junctions considered for memory application as well.
While the ER effect in symmetric FTJ is not as high as in asymmetric junction,
using of the same material for both electrode is an advantage for fabrication process.

The above mentioned mechanisms deform the potential barrier height, thickness 
and shape leading to the ER effect. There is however another mechanism
modifying the barrier in TJ, namely the image forces acting
on electron moving through the barrier. It is known that image forces reduce
the barrier height and its thickness in TJ~[\onlinecite{Lundqvist}]. It is also
known that the strength of image forces depends on the dielectric constant
of the barrier. In FEs the dielectric constant depends on the external parameters
such as the applied voltage, temperature and the direction of FE polarization
(at non-zero bias). Thus, one can control image forces in FE barrier with
external parameters in FTJ and therefore there is one more way to control the
barrier parameters and the tunnelling probability. In the present paper we
investigate the influence of the image forces on the conductance of the FTJ.
We will show that image forces may also produce the ER effect at non-zero bias
in symmetric and asymmetric FTJ. In strongly asymmetric junctions the
image forces can be neglected while in symmetric TJ the presence of image forces
is crucial.
\begin{figure}
\includegraphics[width=1\columnwidth]{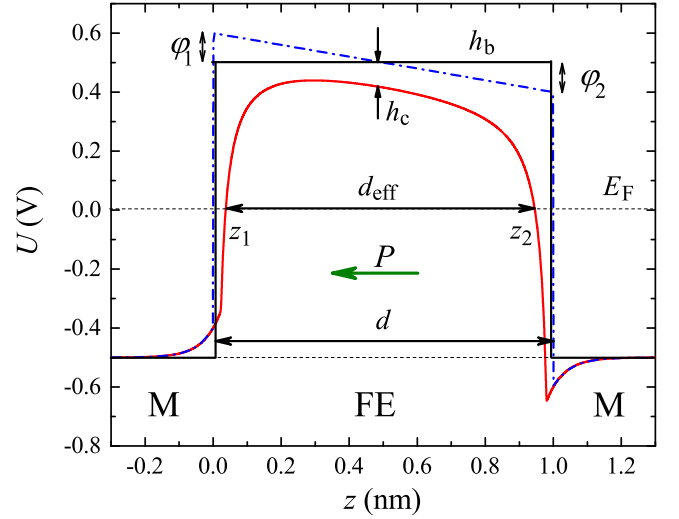}
\caption{(Color online) Potential barrier $U(z)$ (Eq.~(\ref{Eq:BarrierGen}))
as a function of distance for symmetric FTJ at zero bias voltage.
Black like is the potential profile
in the absence of surface charges effect and image forces. Dash-dotted blue line
is the potential corrected due to the surface charges. Red solid line is the
potential profile corrected by both the image forces and FE surface charges.
Notations M and FE stand for metal and ferroelectric, respectively.
FE layer thickness is $d$. $z_{1,2}$ is the position where potential $U(z)$ crosses the
Fermi energy $E_\mathrm F$ (which is the zero energy). $z_{1,2}$ defines
the effective barrier thickness $d_\mathrm{eff}=z_2-z_1$; $P$ denotes the
FE polarization and potentials $\varphi_{1,2}$, $h_{b,c}$ are
introduced in the text.}\label{Fig:Barrier1}
\end{figure}

Recently, the dependence of FTJ conductance on temperature was investigated in asymmetric FTJ~[\onlinecite{Chu2015}]. The effect was related to the variation of polarization (surface charges) with temperature. It exists only below the FE Curie point. Here we show
that similar effect may occur due to the image forces. In FE the dielectric constant strongly varies with temperature leading to variations of image forces strength and thus to the temperature dependence of the FTJ conductance. Since the dielectric constant varies with temperature both below and above the FE Curie point the dependence of the FTJ conductance on temperature should appear in the whole temperature range.

Recently, the influence of image forces were considered and observed in hybrid systems
consisting of FE and thin film of granular metal (GM). Image forces lead to strong dependence
of granular film conductivity on temperature~[\onlinecite{Bel2014GFE1}].
In a field effect transistor with granular channel and the FE placed between the channel and the
gate electrode the image forces lead to the ER effect~[\onlinecite{Bel2014GFE2}]. The dependence
of the granular film conductivity on temperature in FE/GM system was observed in
Refs.~[\onlinecite{Keller2014,Plank2014}]. It was demonstrated that image forces
influence the strength of the Coulomb blockade effect and influence the conductivity of
granular metals. 
In Refs.~[\onlinecite{Bel2014ME1},\onlinecite{Bel2014ME2}] it was theoretically shown 
that the image forces are responsible for coupling between FE substrate 
and magnetic granular film (magneto-electric effect).

It is important that typical FEs have a high dielectric 
constant (of order of 1000). 
Such FEs are not suitable for observation of image forces effects, 
since the strength of image forces is inversely proportional to the FE 
dielectric susceptibility.  FEs with low dielectric constant are more suitable. 
There are a number of low dielectric constant FEs such as hafnium oxide family XHfO$_2$ (where X can be Y,  Co, Zr, Si)~[\onlinecite{Mikolajick2012,Hwang2014,Mikolajick2011}], rare-earth manganites XMnO$_3$ (where X is the rare-earth element)~[\onlinecite{Tokura2004}], colemanite~[\onlinecite{FATUZZO1960}], Li-doped ZnO~[\onlinecite{Liu2003}], etc. There are also numerous organic FEs with low dielectric constant~[\onlinecite{Boer2010,Xiong2012,Tokura2008}]. 
Most FTJ up to date were fabricated with BTO barrier having a very high dielectric constant. 
One can neglect image forces in this type of TJs. However, the low endurance and extremely 
complicated fabrication process restricted applications of oxide FEs so far. 
Organic FEs provide an alternative approach~[\onlinecite{Chu2016}] with low 
cost processing and opportunities to fabricate flexible electronic devices. Such FEs are in 
the track of emerging field of organic electronics. Note that  FTJ with organic FE were 
fabricated and showed significant TER effect recently~[\onlinecite{Chu2016}]. 
Organic FEs mostly have a low dielectric constant and image forces should play a crucial 
role in FTJ with organic FEs. Another important point is that dielectric properties 
of FE materials weaken with decreasing of film thickness~[\onlinecite{Fridkin2006}]. 
FEs with moderate dielectric constant may have a rather weak dielectric response 
as they embedded as few nm thick layer in FTJ.

The paper is organized as follows. We introduce the model and 
calculation procedure in Sec.~\ref{Sec:AnFor}. Analytical estimates of 
influence of image forces and surface charges on the barrier 
parameters are given in Sec.~\ref{Sec:AnEst}. 
In Sec.~\ref{Sec:Res} we present analysis of ER effect due to image forces.

\section{The model}\label{Sec:AnFor}

To study tunnelling currents in FTJ we use the following model.
Consider a FTJ with FE barrier having polarization $P$, dielectric constant $\epsi$ and
thickness $d$ (see Fig.~\ref{Fig:Barrier1}). The polarization is assumed to be uniform
across the barrier and directed perpendicular to the barrier surfaces.
A voltage $V$ is applied to the FTJ. The leads of the FTJ are made of good metals
with the Fermi momentum $k_\mathrm {F1}$ and $k_\mathrm {F2}$, respectively.
We assume that the Fermi energy of the leads is large enough such that the screening lengths
in these metals $\delta_{1,2}$ are small, ($\delta_{1,2}\ll d$). In this case one can
use a simple picture of image forces to describe the correlation effects
inside the insulating barrier. Following Refs.~[\onlinecite{Lundqvist,Tsymbal2005}]
we describe the barrier seen by transport (close to Fermi level) electrons measured in Volts as follows (region $0<z<d$)
\begin{equation}\label{Eq:BarrierGen}
U(z)=h_\mathrm b+\left(\varphi_1-(\varphi_1-\varphi_2)\frac{z}{d}\right)+\frac{0.795ed}{16\pi\eps0\epsi z(d-z)}-V\frac{z}{d},
\end{equation}
with
\begin{equation}\label{Eq:SurfPot}
\begin{split}
&\varphi_1=\frac{dP\delta_1}{\eps0(d+\epsi(\delta_1+\delta_2))},\\
&\varphi_2=-\frac{dP\delta_2}{\eps0(d+\epsi(\delta_1+\delta_2))}.
\end{split}
\end{equation}
Here $e$ is the electron charge, $\eps0$ is the vacuum dielectric constant, $z$ is
the coordinate perpendicular to the layers surfaces; $h_\mathrm b$ defines
the barrier height above the Fermi level of the left lead (which is energy zero in our model)
in the absence of FE polarization, image forces and external voltage.
Potentials $\varphi_{1,2}$ in Eq.~(\ref{Eq:BarrierGen}) are due to formation of
surface charges at the FE/metal interfaces. These charges occur due to
polarization of the FE layer as well as due to screening of polarization
by electrons in metallic leads. The potentials are found 
using Thomas-Fermi approximation with close circuit conditions (see Ref.~[\onlinecite{Tsymbal2005}]).

The third term in Eq.~(\ref{Eq:BarrierGen}) describes
the influence of the image forces. These forces appear due to the interaction
of electron inside the barrier with image charges occurring in metallic leads.
Calculating the image forces potential we consider metallic leads as
ideal, neglecting corrections due to finite screening length.
When calculating potentials $\varphi_{1,2}$ the finite screening length is crucial and can
not be neglected.

The last term in  Eq.~(\ref{Eq:BarrierGen}) describes the effect of the applied voltage.

In our model the Fermi energy of both metals is larger than
potentials $\varphi_{1,2}$ and $V$ 
($|\varphi_{1,2}|+|V|<\hbar^2k^2_\mathrm{F1,2}/(2m_\mathrm e e)$, $m_\mathrm e$ is the electron mass).
This means that potentials created by the FE polarization together with
voltage do not produce the charge depleted layer inside the leads.
This is in contrast to the case of FTJ having at least one lead with small Fermi energy 
considered in numerous papers. In such a FTJ the surface charges turn a metal lead 
with small Fermi level into an
insulator in the vicinity of FE/metal interface leading to the increase of effective barrier width.

In our model the effective barrier thickness can be decreased due to image
forces or in the situation when the potentials $\varphi_{1,2}$ and $V$ exceed
the barrier height ($|\varphi_{1,2}|+|V|>h_\mathrm{b}$).
This situation may easily occur if
the barrier height is less than $1$ eV.

\subsection{FE layer}

We use the following model of FE layer:
below the Curie point the spontaneous FE polarization is a function of
applied voltage and has a hysteresis with the switching voltage $V_\mathrm s$
and the saturation polarization $P_0$. We use the following formula capturing
these peculiarities of FE layer
\begin{equation}\label{Eq:Pol}
P^{\pm}(V)=P_0 \frac{1-e^{-(V\mp V_\mathrm s)/\Delta V_\mathrm s}}{1+e^{-(V \mp V_\mathrm s)/\Delta V_\mathrm s}},
\end{equation}
where ``$+$'' and ``$-$'' correspond to the upper and the
lower hysteresis branch respectively, $\Delta V_\mathrm s$ is the width of
the transition region. For example, the polarization of HfZrO$_2$ is shown in Fig.~\ref{Eq:Pol} and
can be approximately described with the following parameters:
$P_0=30$ $\mu$C/cm$^2$, $V_\mathrm s=d \cdot 10^8$ V (with $d$ being measured in m), and $\Delta V_\mathrm s=0.4 V_\mathrm s$.
We use these values of $V_\mathrm s$ and $\Delta V_\mathrm s$ in all our calculations.
The parameters were obtained by fitting the experimental curves of Ref.~[\onlinecite{Mikolajick2012_1}].
TbMnO$_3$ which also have rather low dielectric constant can be described with the 
following parameters: 
$P_0=7.5~\mu$C/cm$^2$,  $V_\mathrm s=0.7d\cdot 10^8$ V, $\Delta V_\mathrm s=3.5 V_\mathrm s$~[\onlinecite{Noh2006}].

We introduce the dependence of dielectric permittivity on applied voltage below the Curie temperature
\begin{equation}\label{Eq:Diel}
\epsi^{\pm}(V)=\varepsilon_\mathrm{min}+\frac{\Delta\varepsilon}{1+(V \mp V_\mathrm s)^2/\Delta V_\mathrm s^2}.
\end{equation}
This dependence captures the basic features of dielectric
constant behavior as a function of electric field.
The dielectric permittivity has two branches corresponding to two
polarization states. In the vicinity of the switching bias the dielectric permittivity,
$\epsi$ has a peak. Note that sometimes in the literature the following function is used $a/\sqrt{b^2+(V-V_\mathrm s)^2}$, where $a$ and $b$ are fitting parameters. There is no qualitative difference between this formula and Eq.~(\ref{Eq:Diel}) in the range of voltages we study. For higher voltages Eq.~(\ref{Eq:Diel}) gives a finite dielectric constant which is more correct than the 
zero $\epsi$ given by $a/\sqrt{b^2+(V-V_\mathrm s)^2}$. The second order phase transition theory gives  $\epsi(V)$ diverging at $V=V_\mathrm s$, which is also not suitable for description of real systems.

\begin{figure}
\includegraphics[width=1\columnwidth]{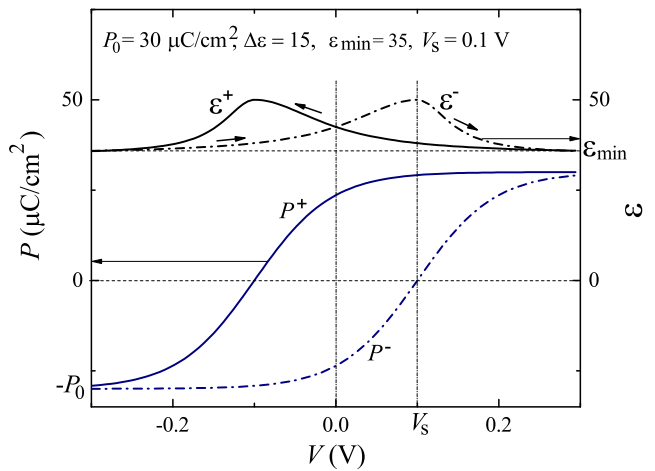}
\caption{(Color online) Polarization (Eq.~(\ref{Eq:Pol})) and
dielectric constant (Eq.~(\ref{Eq:Diel})) shown for
the following parameters: $P_0=30~\mu$C/cm$^2$, $V_\mathrm s=0.1$ V,
$\Delta \epsi= 30$ and $\epsi_\mathrm{min}=15$. Solid lines correspond to the
upper hysteresis branch. Dash-dotted lines stand for the
lower hysteresis branch. The parameters correspond to Hf$_{0.5}$Zr$_{0.5}$O$_2$ FE
(see Ref.~[\onlinecite{Mikolajick2012_1}])}\label{Fig:PolEpsVsV1}
\end{figure}

Not much data are currently available on voltage dependencies of $\epsi(V)$ for 
FEs with low dielectric constants. For example, the dielectric constant of HfZrO$_2$ can be described using the
following parameters: $\varepsilon_\mathrm{min}=35$, $\Delta\varepsilon=15$ (see Fig.~\ref{Fig:PolEpsVsV1}).  

The dielectric constant of TbMnO$_3$ has a much lower variation of 
dielectric constant,  
$\varepsilon_\mathrm{min}=19$, $\Delta\varepsilon=2$~[\onlinecite{Noh2006}]. 
Therefore HfZrO$_2$ is better suited for checking our predictions.

Below we also study the temperature dependence of FTJ conductance 
using experimental data on $\epsi(T)$. We model the temperature dependence 
of FE dielectric constant using the following formula
\begin{equation}\label{Eq:DielT}
\epsi(T)=\left\{\begin{split} &\epsi_\mathrm{min1}^T+\frac{\Delta\epsi_1^T}{\sqrt{(T-T_\mathrm C)^2+\Delta T_1^2}},~~T>T_\mathrm C, \\ &\epsi_\mathrm{min2}^T+\frac{\Delta\epsi_2^T}{\sqrt{(T-T_\mathrm C)^2+\Delta T_2^2}},~~T<T_\mathrm C. \end{split}\right.
\end{equation}
This function allows to capture all peculiarities of $\epsi(T)$ behavior,
namely the finite height peak at $T=T_\mathrm C$ (where $T_\mathrm C$ is the
FE Curie temperature), $1/(T-T_\mathrm C)$ dependence aside the immediate
vicinity of $T=T_\mathrm C$ as well as asymmetry of $\epsi(T)$ curve
with respect to the point $T=T_\mathrm C$. We keep the function continuous at the point $T=T_\mathrm C$.

\subsection{Calculation of resistance}\label{Sec:CurrDef}

We assume that the FE barrier is thin enough and the electron transport 
occurs due to tunnelling. To calculate the electric current across the barrier we
use Simmon's formula~[\onlinecite{Simmons1963,Simmons1963_1}]
\begin{equation}\label{Eq:Curr}
J=J_0(\overline U(h_\mathrm b) e^{-A\sqrt{\overline U(h_\mathrm b)}}-\overline U(h_\mathrm b+V)e^{-A\sqrt{\overline U(h_\mathrm b+V)}} ),
\end{equation}
where
\begin{equation}\label{Eq:AvBar}
\sqrt{\overline U}=\frac{1}{d_\mathrm{eff}}\int_{z_1}^{z_2}\sqrt{U(z)}dz,
\end{equation}
the parameter $A=\beta d_\mathrm{eff}\sqrt{2m_\mathrm e e/\hbar^2}$ and $J_0=(e^2/\hbar\beta d_\mathrm{eff}^2)$.
The integration in Eq.~(\ref{Eq:AvBar}) is performed over the region where $U(z)>0$
which can differ from the region [$0,d$] due to surface charges and image forces effects.
The coordinates where $U(z)=0$ are denoted as $z_1$ and $z_2$
and $d_\mathrm{eff}=z_2-z_1$ is the effective barrier thickness.
The constant $\beta$ is of order of 1. 
Equation~(\ref{Eq:Curr}) is just the difference between currents 
created by electrons in left (the first term) and right (the second term) leads. 
Since the Fermi level in the right lead is biased by the applied voltage, 
the potential barrier seen by electrons in this lead is higher (lower) by $V$ 
for positive (negative) voltage. Therefore, the average potential in the 
second term of Eq.~(\ref{Eq:Curr}) is calculated with $h_\mathrm b$ replaced 
by $h_\mathrm b+V$.

Below we find the TJ resistance $R^\pm (V)$ and the conductance $G^\pm (V)$
\begin{equation}\label{Eq:Res}
R^\pm (V)=\frac{V}{J^\pm (V)},  \, \, \, \, ~G^\pm (V)=\frac{1}{R^\pm (V)}.
\end{equation}
We use the superscript ``$+$'' (``-'') to describe conductance and 
resistance corresponding to the upper (lower) hysteresis branch of FE layer. 

Also we introduce the conductance $G_{\mathrm P}^{\pm}$ for
TJ in the absence of image forces and the
conductance $G^{\pm}_{\epsi}$ which neglects the surface charges effect.

The ER effect due to both polarization and image forces is given by the ratio, $ER=G^{+}/G^-$.
We use the subscripts $\mathrm P$ or $\epsi$ to denote the 
ER effect caused by the surface charges ($ER_\mathrm P$)
or the image forces ($ER_{\epsi}$), respectively. 
To calculate $ER_\mathrm P$ and $ER_{\epsi}$ we neglect the third and the second term in Eq.~(\ref{Eq:BarrierGen}), respectively.

\section{Tunnelling barrier average height and thickness. Analytical estimates}\label{Sec:AnEst}

In this section we compare the influence of image forces and surface charges on the 
average TJ barrier parameters.

\subsection{General remarks}

Since the dielectric constant $\epsi$ at zero bias
is the same for both branches ($\epsi^+|_{V=0} =\epsi^-|_{V=0}$),
the image forces do not lead to the dependence of the
linear TJ resistance on the FE polarization state ($G^+_{\epsi}|_{V=0}=G^-_{\epsi}|_{V=0}$)
even for asymmetric FTJ with $\delta_1\ne \delta_2$.  In contrast, the FE surface
charges produce the ER effect in asymmetric FTJ even at zero bias, $G^+_{\mathrm P}|_{V=0}\ne G^-_{\mathrm P}|_{V=0}$. 
At finite zero-bias both mechanisms lead to the ER effect.

First, we compare the influence of the image
force mechanism and the surface charge mechanism for the
case of symmetric TJ. Second, we discuss the case of asymmetric TJ.

\subsection{Influence of the image forces}

Image forces reduce both the barrier height and the thickness.
The characteristic potential associated with image forces in TJ is given by the expression
\begin{equation}
h_\mathrm c=\frac{0.795e^2}{4\pi\eps0\epsi d}\frac{1}{e}.
\end{equation}
This is the reduction of the initial potential barrier height (see Eq.~(\ref{Eq:BarrierGen})) at
the symmetry point ($z=d/2$, see Fig.~\ref{Fig:Barrier1}) at zero bias.
For $\epsi=5$ and $d=1$ nm we have $h_\mathrm c=0.25$ V. The
ratio of $h_\mathrm c$ and $h_\mathrm b$ defines the effective barrier
thickness $d_\mathrm{eff}$ as follows
\begin{equation}
d_\mathrm{eff}=d\sqrt{1-\frac{h_\mathrm c}{h_\mathrm b}}.
\end{equation}
One can see that the influence of the image forces on the height
and the thickness increases with decreasing of FE dielectric constant and the
barrier thickness. The barrier thickness reduction exceeds 10\% only for $\epsi<10$ even for the smallest
feasible $d$. Therefore we can always treat it as a small perturbation
and approximate $d_\mathrm{eff}\approx d(1-h_\mathrm c/(2h_\mathrm b))$.
In Sec.~\ref{Sec:CurrDef} we introduced the average barrier height as follows,
$\overline U=(\int \sqrt{U(z)}dz/d_\mathrm{eff})^2$. We use this expression
in our numerical calculations. For analytical consideration of the influence of
image forces it is enough to use a simpler expression,
$\overline{U}=\int U(z)dz/d_\mathrm{eff}$, where integration is over the
region [$z_1,z_2$] and the term with $\varphi_{1,2}$ is neglected.
The tunnelling probability is defined by the product of potential barrier and its average height
\begin{equation}\label{Eq:AvBarrierImF}
d_\mathrm{eff}\sqrt{\overline{U}}\approx d\sqrt{h_\mathrm b}\left(1-\frac{h_\mathrm c}{2h_\mathrm b}\left(1+\frac{1}{2}\mathrm{ln}\frac{h_\mathrm b}{4h_\mathrm c}\right)\right).
\end{equation}
Corrections due to the image forces are defined by the ratio $h_\mathrm c/h_\mathrm b$.
These estimates show that correction to the resistance due to image forces
increases with decreasing of FE dielectric constant $\epsi$, but
is independent of the barrier thickness $d$ since the common factor $d$
is compensated by $d^{-1}$ in the factor $h_\mathrm c$. Decreasing the barrier
height $h_\mathrm b$ also increases the influence of the image forces.

\subsection{Influence of FE surface charges}

First we consider the symmetric case, $\delta_1=\delta_2$, in which 
the potential $\varphi_1-(\varphi_1-\varphi_2)\frac{z}{d}$ is
an odd function of $z-d/2$ and gives a zero contribution to
the average potential, $\int U(z)dz/d_\mathrm{eff}$.  The effect of
surface charges appears only due to the fact that $U(z)$ enters the
tunnelling probability in a non-linear way. Therefore in the case of surface
charges one should use the average potential calculated in
Eq.~(\ref{Eq:AvBar}) to estimate its
influence. Here we assume that potentials $\varphi_{1,2}$ are
smaller than the Fermi energy and the surface charges do not change the barrier thickness. We have
\begin{equation}\label{Eq:AvBarrierPol}
d\sqrt{\overline U}=\frac{-2d}{3(2\varphi_1+V)}\left(\sqrt{h_\mathrm b-\varphi_1-V}^3-\sqrt{h_\mathrm b+\varphi_1}^3\right).
\end{equation}
For symmetric TJ we have $\varphi_2=-\varphi_1$. For small $\varphi_1$ and $V$ ($\varphi_1,V\ll h_\mathrm b$) the above
expression can be written as
\begin{equation}\label{Eq:AvBarPol1}
d\sqrt{\overline U}=d\sqrt{h_\mathrm b}\left(1-\frac{V}{h_\mathrm b}-\frac{1}{24}\frac{(\varphi_1+V)^3+\varphi_1^3}{h_\mathrm b^2(2\varphi_1+V)}\right).
\end{equation}

The ratio $\varphi_1/h_\mathrm b$ defines the contribution of
surface charges to the effective barrier height.
One can see that the surface charges effect decreases faster
with increasing of $h_\mathrm b$ (as $h^{-3/2}_\mathrm b$) in comparison with the
effect of image forces. It is also important that
the last term in the brackets may change its sign.

We estimate $\varphi_1\approx dP/(2\epsi\eps0)$ for $d<2\epsi\delta_1$. For $\varepsilon=50$, $P=20~\mu$C/cm$^2$, $d=1$ nm and $\delta_1=0.05$ nm we find $\varphi_1\approx 0.25$ V. The potential $\varphi_1$ can be
increased with increasing thickness $d$. Also, there are many materials with polarization larger
than $20~\mu$C/cm$^2$. Thus, in contrast to the case of image forces the influence of surface
charges can be made very strong. The influence of surface charges decreases with
decreasing of polarization $P$, and thickness $d$. The dielectric permittivity
enters this mechanism in the same way as in the mechanisms
based on image forces. However, as we show below,
the dependencies of $ER_{\epsi}$ and $ER_\mathrm P$ on $\varepsilon$ are different. 
The influence of dielectric permittivity voltage dependence on the surface 
charges ER effect was studied in Ref.~[\onlinecite{Xiao2011}].
The potential $\varphi_1$ decreases with $\delta_1$ for $d>2\epsi\delta_1$,
however this regime is difficult to reach for
thickness $d$ of about 1 nm.

\subsection{Comparing image forces and surface charge contributions to ER effect}

The ER effect is defined by the change of the effective barrier height and thickness under the change of FE state
(approximately $ER\sim\mathrm{exp}(d_\mathrm{eff}\sqrt{\overline{U}}|_{P^+}-d_\mathrm{eff}\sqrt{\overline{U}}|_{P^-})$).
For symmetric barrier the expression in the exponent is non-zero only for finite bias.
The polarization switching leads to the change of dielectric
constant ($\epsi^+|_{V\ne 0}\ne \epsi^-|_{V\ne 0}$). The variation of
the barrier due to the image forces is given by
\begin{equation}\label{Eq:EstImF}
d_\mathrm{eff}\sqrt{\overline{U}}|_{P^+}-d_\mathrm{eff}\sqrt{\overline{U}}|_{P^-}\sim d\sqrt{h_\mathrm b} \frac{h^0_\mathrm c}{h_\mathrm b}\frac{\epsi^+-\epsi^-}{\epsi^+ \epsi^-},
\end{equation}
where $h^0_{\mathrm c}=h_{\mathrm c} |_{\epsi=1}$. The barrier variation due to the
surface charges has the form
\begin{equation}\label{Eq:EstP}
d_\mathrm{eff}\sqrt{\overline{U}}|_{P^+}-d_\mathrm{eff}\sqrt{\overline{U}}|_{P^-}\sim \frac{d^2\sqrt{h_\mathrm b}PV}{2\epsi \epsi_0 h^2_\mathrm b}.
\end{equation}
Comparing these expressions we can estimate the ratio of two mechanisms contributing
to the ER effect $(\delta \epsi/\epsi)(e/(d^2 P))$ (we use $V\sim h_\mathrm b$).
In this ratio $\delta \epsi$ is the difference of dielectric constants for positive
and negative polarization, $\delta \epsi = \epsi^+-\epsi^-$. If this ratio is larger
than 1 then the image forces define the ER effect, otherwise the surface
charges are more important. For example, for $d=1$ nm and $(\delta \epsi/\epsi)\approx 30$ \% the
image forces mechanism is more pronounced for $P<5~\mu$C/cm$^2$.

\subsection{Asymmetric TJ}

For asymmetric TJ the resistance at zero bias depends on the direction of polarization.
This effect appears due to the surface charges of the FE layer in combination with
asymmetric screening of these charges. Here we compare this ER effect at zero bias
with ER at $V=V_\mathrm s$ due to image forces.
\begin{figure}
\includegraphics[width=1\columnwidth]{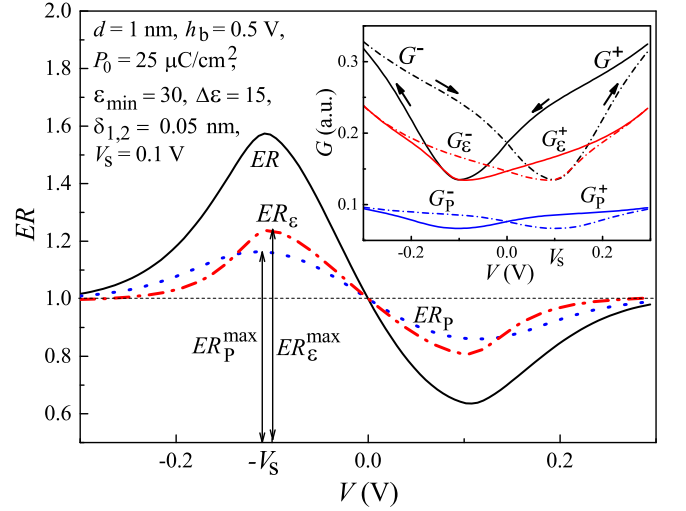}
\caption{(Color online) Electro-resistance effect (main plot) and
conductance (inset) as a function of applied voltage for the
following system parameters: $d=1$ nm, $h_\mathrm b=0.5$ eV,
$P_0=30~\mu$C/cm$^2$, $\delta_1=\delta_2=0.05$ nm, $V_\mathrm s=0.1$ V,
$\Delta \epsi = 15$ and $\epsi_\mathrm{min}=30$. Solid line corresponds
to the total ER effect including both mechanisms related to surface
charges and image forces.
Dotted line shows the ER effect due to surface charges $ER_\mathrm P$.
Dashed-dotted line shows the ER effect due to the image
forces $ER_{\epsi}$. $ER^\mathrm{max}_{\epsi , \mathrm P}$ is the
value of $ER_{\epsi}$ and $ER_\mathrm P$ at $V=-V_\mathrm s$. Inset:
solid lines correspond to the upper hysteresis branch.
Dash-dotted lines stand for lower hysteresis branch.
Arrows show the path of hysteresis loop.}\label{Fig:ERvsV1}
\end{figure}

For asymmetric TJ the average potential (neglecting image forces)
is given by the expression, $\overline U\approx h_\mathrm b+\frac{dP(\delta_1-\delta_2)}{\eps0\epsi(\delta_1+\delta_2)}$
and the relative change of $d\sqrt{\overline U}$ is given by
$\frac{dP(\delta_1-\delta_2)}{\eps0\epsi(\delta_1+\delta_2)h_\mathrm b}$.
This expression does not take into account a variation of barrier thickness
appearing for leads with Fermi energy smaller than $\varphi_{1,2}$.
Comparing the expression with the relative barrier change
due to image forces we can write the
ratio $(\frac{e}{d^2P})(\frac{\Delta \epsi}{\epsi}/\frac{\Delta \delta}{\delta})$.
We can neglect the image forces in asymmetric TJ
if this ratio is larger than 1. This is the case when BTO or PZT FE is used in
asymmetric TJ made of Pt and LSMO metals. We note that to create the
TJ with $\Delta \delta/\delta\approx 10$\% one has to use the metals with Fermi
level difference of about 40\%, since $\delta\sim E^{1/4}_\mathrm F$.

\section{Electro-resistance effect in FTJ}\label{Sec:Res}

In this section we compare the ER effect for symmetric and asymmetric FTJ
appearing due to image forces and due to surface charges. We discuss the behavior of the
ER effect on system parameters such as barrier height and thickness,
saturation polarization and dielectric constant.
We calculate all curves using Eqs.~(\ref{Eq:Curr}), (\ref{Eq:AvBar}) and (\ref{Eq:Res}).

\subsection{Symmetric FTJ}

The inset in Fig.~\ref{Fig:ERvsV1} shows the behavior of FTJ conductance for
symmetric barrier as a function of applied bias. The curves have a hysteresis
character originating from the hysteresis of polarization $P$ and the
dielectric constant $\epsi$. For symmetric TJ with $\delta_1=\delta_2$ the conductance
at zero bias is the same for both hysteresis branches.
At non-zero bias the symmetry of the TJ is broken due to external field
leading to difference in conductance for different hysteresis branches.
The chosen parameters correspond to Hf$_{0.5}$Zr$_{0.5}$O$_2$ FE material.
\begin{figure*}[t]
	\begin{center}
		\includegraphics[width=1\textwidth, keepaspectratio]{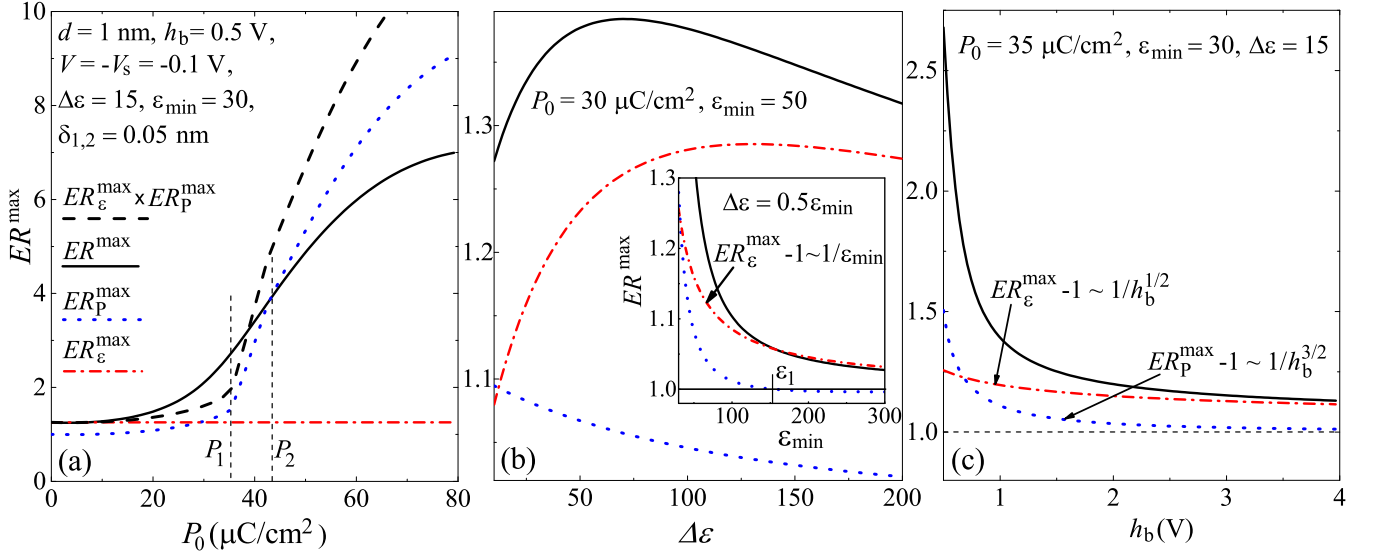}
		\caption{(Color online)  Maximum value of electro-resistance
		effect $ER^\mathrm{max}$ as a function of (a) saturation
		polarization $P_0$, (b) dielectric constant variation
		$\epsi_{\mathrm{min}}$ and $\Delta\epsi$, and (c) barrier height $h_\mathrm b$.
		The following parameters are used for all three plots:
		$d=1$ nm, $\delta_{1,2}=0.05$ nm, $V=-V_\mathrm s=-0.1 $ V. Solid lines correspond
		to the total ER effect including effects of surface charges
		and image forces $ER^\mathrm{max}$. Dotted line shows the ER effect due to
		surface charges $ER^\mathrm{max}_\mathrm P$. Dashed-dotted line is the ER effect
		produced by the image forces $E^\mathrm{max}_{\epsi}$.
(a) Dashed line shows $ER^\mathrm{max}_{\epsi}\times ER^\mathrm{max}_{\mathrm P}$. The barrier
height $h_\mathrm b=0.5$ V,  $\Delta \epsi = 15$
and $\epsi_\mathrm{min}=30$. $P_{1,2}$ denotes polarization where
$ER^\mathrm{max}_\mathrm P$ has a derivative gap. (b) $P_0=30~\mu$C/cm$^2$,
$h_\mathrm b=0.5$ V, $\epsi_\mathrm{min}=50$ for the main graph and
 $\Delta \epsi = 0.5\epsi_\mathrm{min}$ for the inset. The ER effect due to surface charges
 disappears at $\epsi=\epsi_1$. (c) $P_0=35~\mu$C/cm$^2$, $\Delta \epsi = 15$ and $\epsi_\mathrm{min}=30$.}
		\label{Fig:ER_vs_PEH1}
	\end{center}
\end{figure*}

There are three pairs of curves in the inset in Fig.~\ref{Fig:ERvsV1}.
Black lines show the conductance $G$ calculated by taking into account both
the surface charge and the image forces effects.
Blue lines show the conductance calculated neglecting the image forces,
$G_\mathrm P$. The conductance $G_\mathrm P$ has its minimum when polarization
switching occurs ($V=-V_\mathrm s$ for the upper branch and $V=V_\mathrm s$ for the
lower branch).
The minimum can be understood as follows:
Surface charges produce the electric field inside the barrier leading
to the linear slope of the potential $U(z)$ (see Fig.~\ref{Fig:Barrier1}).
The applied bias also creates the electric field inside the barrier. 
According to Eq.~(\ref{Eq:AvBarrierPol}) the stronger the total field the lower the TJ resistance.
Depending on the sign of polarization and voltage these two fields
can enhance (co-directed) or counteract each other (counter-directed).
Consider the positive bias ($V>0$). For upper branch the fields are co-directed and
the conductance grows (see Eq.~(\ref{Eq:AvBarrierPol})).
For lower branch the fields are counter-directed decreasing the
conductance. When the bias reaches the switching voltage $V_\mathrm s$ the
FE polarization changes its sign and both fields (due to bias and due to polarization)
become co-directed for the lower branch. Further bias increase leads
to the increase of the lower branch conductance.

Red lines in the inset in Fig.~\ref{Fig:ERvsV1} are for conductance $G_{\epsi}$.
If one moves left along the upper branch (decreasing voltage starting
with large positive bias) the dielectric constant reaches its maximum value
at negative bias (at $V=-V_\mathrm s$). The maximum dielectric constant weakens
the image forces doing the barrier higher. Thus, the conductance decreases
in the vicinity of $V=-V_\mathrm s$ for the upper branch
($V=V_\mathrm s$ for the lower branch).
Thus, the change of conductance due to surface charges and image forces
behaves similarly. However, the average
(over the whole voltage region) conductance, $G_{\epsi}$, exceeds the average $G_\mathrm P$,
meaning that image forces influence the conductance
much stronger than the surface charges for given parameters.

Typical dependencies of ER effect on the applied voltage in symmetric FTJ are
shown in Fig.~\ref{Fig:ERvsV1}. The black solid line shows the ER effect
calculated by taking into account both the image forces and the
surface charges ($ER$), the blue dotted line is for the ER effect calculated by
taking into account only the surface charges ($ER_\mathrm P$), the
red dash-dotted line corresponds to the ER effect due to image
forces ($ER_{\epsi}$). The parameters for which the curves were
calculated correspond to Hf$_{0.5}$Zr$_{0.5}$O$_2$ FE. Due to the symmetry
of the hysteresis loop in Eqs.~(\ref{Eq:Pol}) and~(\ref{Eq:Diel}) the
ER effect obeys the relation $ER(V)=ER(-V)^{-1}$. At zero bias the
conductance does not depend on the FE state and the ER effect is absent ($ER=1$).
At high voltage the FE state is the same for both branches and the ER effect is absent.
The ER effect reaches its maximum value at switching voltage. We denote
it $ER^{\mathrm{max}}_{\epsi}$ for ER effect due to image
forces, $ER^{\mathrm{max}}_\mathrm P$ for ER effect due to the
surface charges, and $ER^{\mathrm{max}}$ for ER effect including
both mechanisms.  One can see that for given parameters the image
forces produce stronger ER effect than the surface charges.
This is in agreement with our analytical estimates showing that
image forces are important for 1 nm thick FTJ.

Figure~\ref{Fig:ER_vs_PEH1} shows the dependence of the maximum value of
the ER effect ($ER^\mathrm{max}_{\mathrm P}$, $ER^\mathrm{max}_{\epsi}$ and $ER^\mathrm{max}$)
on the parameters of FE barrier (saturation polarization $P_0$,
dielectric constant variation $\epsi_\mathrm{min}$ and $\Delta\epsi$
and the barrier height $h_\mathrm b$). First, consider the left panel in the figure.
When calculating the dependencies on polarization (Fig.~\ref{Fig:ER_vs_PEH1}(a))
we fixed the switching voltage and the dielectric constant for the upper and
the lower branch. The dotted line corresponds to the maximum ER effect due
to surface charges. At zero $P_0$ this mechanism does not lead to the ER effect. The
ER effect grows with increasing $P_0$. The growth regime changes at
points $P_{1,2}$. For small polarization and $\varphi_{1,2}<h_\mathrm b$
the surface charges change the average barrier height
according to Eq.~(\ref{Eq:AvBarrierPol}).
For $\varphi_{1,2}>h_\mathrm b$ the surface charges change the
effective barrier thickness (this effect is absent
in Eq.~(\ref{Eq:AvBarrierPol})). This leads to discontinuity
of the derivative of $ER^\mathrm{max}_{\mathrm P}$. At $P_0=P_1$ the
condition $h_\mathrm b-V_s+\varphi_2(P)=0$ is satisfied for
the lower polarization branch. At $P_0=P_2$ the same condition
is fulfilled for the upper branch.

The red dash-dotted line shows the ER effect due to the image forces.
It does not depend on the saturation polarization. Comparing $ER^\mathrm{max}_{\epsi}$
and $ER^\mathrm{max}_{\mathrm P}$ one can see that for small polarization the
image forces effect, $ER^\mathrm{max}_{\epsi}$, exceeds the surface charge effect,
$ER^\mathrm{max}_{\mathrm P}$. For large saturation polarization the situation is the opposite.

Interestingly that the total ER effect $ER^\mathrm{max}$ is not
just a product of $ER^\mathrm{max}_{\epsi}$ and $ER^\mathrm{max}_{\mathrm P}$ shown
with black dashed line. At low saturation polarization the total
ER effect exceeds $ER^\mathrm{max}_{\epsi}\times ER^\mathrm{max}_{\mathrm P}$
while at large polarization one has,
$ER^\mathrm{max}<ER^\mathrm{max}_{\epsi}\times ER^\mathrm{max}_{\mathrm P}$.

Figure~\ref{Fig:ER_vs_PEH1}(b) shows the dependence of the maximum ER
effect on the variation of dielectric constant $\Delta\epsi$ (main graph)
and on $\epsi_{\mathrm{min}}$ (inset). A variation of dielectric constant is the
reason for ER effect due to image forces according to our analytical
estimates, Eq.~(\ref{Eq:EstImF}). Therefore, $ER^\mathrm{max}_{\epsi}$ grows
with increasing of $\Delta\epsi$ reaching its maximum value when
variation of dielectric constant becomes of the same order as the
average dielectric constant ($\epsi_\mathrm{min}+\Delta\epsi/2$).
Further increase of dielectric constant variation leads to the decrease of
the ER effect. According to Eq.~(\ref{Eq:EstImF})
the ER effect behaves as $(\epsi^+-\epsi^-)/(\epsi^+\epsi^-)$. The numerator of this
expression grows with $\Delta\epsi$. But the denominator grows too due to
the finite width of the transition region, $\Delta V_\mathrm s$.
According to Eq.~(\ref{Eq:Diel}) increasing of $\Delta\epsi$ leads to the increase
of both $\epsi^+$ and $\epsi^-$. Thus at a certain value $\Delta\epsi$ the ER effect
starts decreasing. Reducing the width of the transition region, $\Delta V_\mathrm s$,
one can increase the ER effect due to the image forces.

In Fig.~\ref{Fig:ER_vs_PEH1}(b) the value of $ER^\mathrm{max}_{\epsi}$ does not exceed several
	tens of percent. Generally, there is no restriction on the value of ER effect.
	One can expect the magnitude of the effect of order of $\Delta\epsi/\epsi_\mathrm{min}$
	for small width of the transition region $\Delta V_\mathrm s$.
	For $\epsi_\mathrm{min}=10$ and $\Delta\epsi=50$ the
	magnitude of $ER^\mathrm{max}_{\epsi}$ can be as high as 7 (the ER effect
	due to image forces is about 700\%) if $V_\mathrm s\to 0$.

The contribution of surface charges simply decreases with
increasing of $\Delta\epsi$. This effect is related to finite width of the transition region
$\Delta V_\mathrm s$.
At $V=-V_\mathrm s$ one has $P^+|_{V=-V_\mathrm s}=0$ and $\varphi_1|_{V=-V_\mathrm s}=0$
for the upper branch. Therefore, only $\epsi^-$ and $P^-$ enter $ER_\mathrm P$. For
$\Delta V_\mathrm s=0$ we have $\epsi^-|_{V=-V_\mathrm s}=\epsi_\mathrm{min}$ and $\Delta \epsi$ 
does not influence $ER_\mathrm P$. 
For finite $\Delta V_\mathrm s$ we have $\epsi^-|_{V=-V_\mathrm s}>\epsi_\mathrm{min}$ and it
grows with increasing of $\Delta \epsi$ leading to the decrease of the ER effect due to surface charges.

The inset in Fig.~\ref{Fig:ER_vs_PEH1}(b) shows the dependence of maximum ER effect
on the minimum dielectric permittivity $\epsi_\mathrm{min}$ for variation of
the dielectric constant $\Delta\epsi$ scaled with $\epsi_\mathrm{min}$ ($\Delta\epsi=0.5\epsi_\mathrm{min}$).
The maximum value of ER effect due to the image forces decays
as $1/\epsi_\mathrm{min}$, in agreement with Eq.~(\ref{Eq:EstImF}). 
The ER effect due to surface charges decays much faster and even changes sign
at the point $\epsi_\mathrm{min}=\epsi_1$. 
For $\epsi_{\mathrm{min}}<\epsi_1$ the potentials $\phi_{1,2}$ are mostly governed by 
dependence $P(V)$ and for $\epsi_{\mathrm{min}}>\epsi_1$ variation 
of  $\epsi$ with $V$ becomes more important. Figure~\ref{Fig:ERvsV1} shows the case with $\epsi_\mathrm{min}<\epsi_1$. In the opposite limit the dependence of the ER effect due to surface charges on voltage, $ER_\mathrm P(V)$, is a reflected version of Fig.~\ref{Fig:ERvsV1}
with respect to the point $V=0$.

Absence of the ER effect at $\epsi_{\mathrm{min}}=\epsi_1$ can be understood using
Eq.~(\ref{Eq:AvBarPol1}). We calculate $ER^\mathrm{max}$ at $V=-V_\mathrm s$.
At this point the upper branch has zero polarization and the lower branch has
the polarization $\approx -P_0$. The conductance for the upper branch is
defined by the quantity $d_{\mathrm{eff}}\sqrt{\overline U}=d\sqrt{h_\mathrm b}(1-V/h_\mathrm b - (1/24)(V^2/h_\mathrm b^2))$.
The potential profile has an upward tilt due to the bias.
The conductance for the lower branch depends on $\varphi_1\ne 0$.
The electric field and the field due to
the surface charges inside the barrier are counter-directed.
For small absolute value of voltage the potential profile $U(z)$ has the
downward tilt larger than the tilt of the potential profile of the
upper branch. Increasing the bias absolute
value we decrease the tilt of $U(z)$ for lower branch and increase the tilt of $U(z)$ for upper branch.
The point $\epsi_1$ (see inset in Fig.~\ref{Fig:ER_vs_PEH1}(b))
is given by the equation $\varphi_1(\epsi_1)=-V_\mathrm s$. When this condition
is satisfied the potential profiles for the upper and the lower branches have
exactly the opposite tilt and the value of $d_{\mathrm{eff}}\sqrt{\overline U}$ is the same for both branches.

The inset in Fig.~\ref{Fig:ER_vs_PEH1}(b) shows that the ER effect due to image forces
exceeds the surface charges effect in a wide range of dielectric constants for symmetric FTJ.
\begin{figure}
	\includegraphics[width=1\columnwidth]{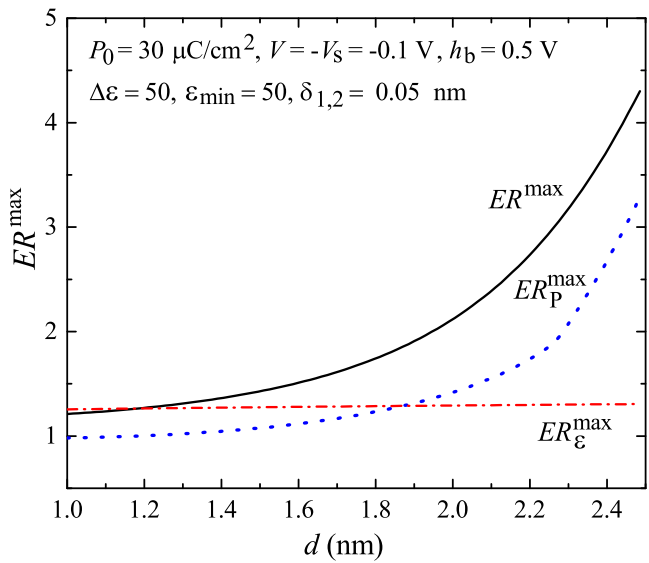}
	\caption{(Color online) Maximum value of electro-resistance
	effect $ER^\mathrm{max}$ (at $V=-V_\mathrm s$) as a function
	barrier thickness $d$  for the following system
	parameters: $\delta_1=\delta_2=0.5$ nm, $P_0=30~\mu$C/cm$^2$,
	$V_\mathrm s=0.1$ V, and $h_\mathrm{b}=0.5$ V, $\epsi_{\mathrm{min}}=50$, $\Delta\epsi = 50$.
	Solid lines correspond to the total ER effect including effects of surface charges
	and image forces. Dotted line shows the ER effect due to the
	surface charges, $ER_\mathrm P$. Dashed-dotted line is the ER effect
	produced by the image forces $ER_{\epsi}$.}\label{Fig:ERvsD1}
\end{figure}

Both contributions to the total ER effect $ER^\mathrm{max}_{\epsi}$ and
$ER^\mathrm{max}_{\mathrm P}$ depend on the barrier height. These dependencies
are shown in Fig.~\ref{Fig:ER_vs_PEH1}(c). One can see that
increasing the barrier height increases the importance of image forces.
For low barrier, $ER^\mathrm{max}_{\mathrm P}$ exceeds $ER^\mathrm{max}_{\epsi}$,
while for high barrier the situation is the opposite. This coincides with
analytical estimates. Equation~(\ref{Eq:EstImF}) shows that
corrections due to image forces to the average barrier height
multiplied by thickness behave as $dh_\mathrm c/\sqrt{h_\mathrm b}$. The magnitude of
$ER^\mathrm{max}_{\epsi}$ behaves similarly. The effect of surface charges
decays as $d\varphi_1V/\sqrt{h_\mathrm b}^3$ according to Eq.~(\ref{Eq:EstP}).
This difference in the behaviour appears due to the fact that image forces
produce the correction even in average potential while the surface
charges give the zero correction to the average potential. The surface charges
contribute to the conductance only if one takes into account the
fact that the tunnelling probability is a function of average of square root of the barrier.

Equation~(\ref{Eq:AvBarPol1}) shows that surface charges contribution
grows with increasing of the screening length. Note that for large
enough screening length, $\delta\sim d$ the approach used in the
manuscript is not valid. The contribution due to the image forces
does not depend on the screening length in our model for $\delta\ll d$.

In the previous section we mentioned that the image forces contribution
becomes less important with increasing of the barrier thickness. This is shown
in Fig.~\ref{Fig:ERvsD1}. One can see that the value $ER^\mathrm{max}_\mathrm P$
grows rapidly with thickness, while $ER^\mathrm{max}_{\epsi}$ \textbf{is} almost independent of $d$.

To summarize this section, we show that the contribution due to image forces
to the ER effect exceeds the surface charge contribution for small barrier thickness
and high barrier height, at small polarization and high variation of dielectric
constant. Increasing the average dielectric constant increases the importance
of image forces contribution reducing the contribution due to surface charges.

\subsection{Asymmetric FTJ}

\begin{figure}
\includegraphics[width=1\columnwidth]{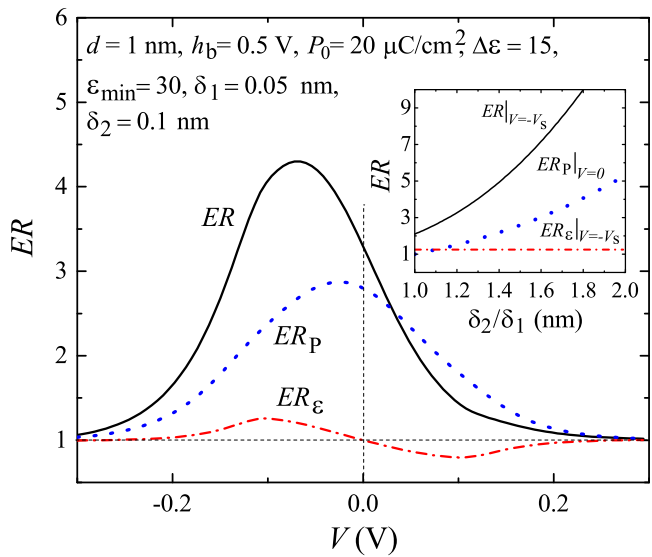}
\caption{(Color online) Electro-resistance (ER) effect as a function of
applied bias for asymmetric FTJ with $\delta_1=0.5$ nm,
$\delta_2=0.1$ nm, $d=1$ nm, $P_0=20~\mu$C/cm$^2$, $V_\mathrm s=0.1$ V,
and $h_\mathrm{b}=0.5$ V, $\epsi_{\mathrm{min}}=30$, $\Delta\epsi = 15$.
Solid lines correspond to the total ER effect including effects of surface charges
and image forces. Dotted line shows the ER effect due to the surface
charges $ER_\mathrm P$. Dashed-dotted line is the ER effect produced by
the image forces, $ER_{\epsi}$. The inset shows the
dependence of the ER effect on the
ratio $\delta_2/\delta_1$ at $\delta_1=0.05$ nm for the same
parameters as in the main plot.}\label{Fig:ERvsDelta1asym}
\end{figure}

For asymmetric FTJ the surface charges produce the ER effect even
at zero bias while image forces do not lead to the ER effect in this case
(see Fig.~\ref{Fig:ERvsDelta1asym}). Therefore, comparison of ER
effect at $V=-V_\mathrm s$ is not a correct way to proceed. Here we compare the
ER effect due to image forces at $V=-V_\mathrm s$ ($E^\mathrm{max}_{\epsi}$)
with the ER effect due to surface charges at zero bias voltage
(see inset in Fig.~\ref{Fig:ERvsDelta1asym}). One can see
that the effect due to surface charges grows rapidly with increasing
the asymmetry and exceeds the ER effect due to image forces.
Even 10\% difference in the screening length produces the ER effect at
zero bias of the same magnitude as the ER effect due to image
forces at $V=-V_\mathrm s$. Note however, that 10\% difference in the
screening length corresponds to 50\% difference in the Fermi
level of the material and $\delta_2/\delta_1=2$ corresponds to
16 times difference of the Fermi levels. Thus, the leads should be made out of
essentially different materials to produce a strong ER effect at zero bias.

\subsection{Temperature dependence of conductance of FTJ}

The important peculiarity of the image forces contribution
to the conductance is related to the fact that it does not
vanish above the ferroelectric Curie point $T_\mathrm C$ while
the surface charges are zero in this temperature region. The dielectric
permittivity of FE strongly depends on temperature above (and below) $T_\mathrm C$.
This leads to a strong dependence of the FTJ conductance on temperature
above and below the phase transition point. Such a dependence above $T_\mathrm C$
occurs only due to the image forces. Using Eq.~(\ref{Eq:AvBarrierImF}) one can
estimate the temperature coefficient of resistance $TCR=-(1/J)(dJ/dT)$ for symmetric FTJ
as follows
\begin{equation}\label{Eq:TCR}
TCR\approx 2\frac{\partial \left(d_\mathrm{eff}\sqrt{\frac{2m_\mathrm e e\overline U}{\hbar^2}}\right)}{\partial T}\approx -d\sqrt{\frac{2m_\mathrm e eh_\mathrm b}{\hbar^2}}\frac{h_\mathrm c}{h_\mathrm b}\frac{1}{\epsi}\frac{\partial \epsi}{\partial T}.
\end{equation}
This quantity is independent of the barrier thickness $d$ and
decays with increasing the barrier height $h_\mathrm b$ and the
average dielectric constant.

So far FTJ were made with only few FE materials: most FTJs have BTO FE with
rather high dielectric constant ($\epsi \sim$ 1000). The temperature dependence of
conductance in symmetric FTJ with BTO due to image forces is very weak,
see Eq.~(\ref{Eq:TCR}). In a recent paper, however the TCR of asymmetric FTJ
with BTO was reported. The dependence of conductance on temperature appears
due to surface charges produced by the FE layer. The dependence of $G(T)$
on temperature occurs below the
FE Curie point and the TCR of order 3.8\% was reported in this system.
\begin{figure}
\includegraphics[width=1\columnwidth]{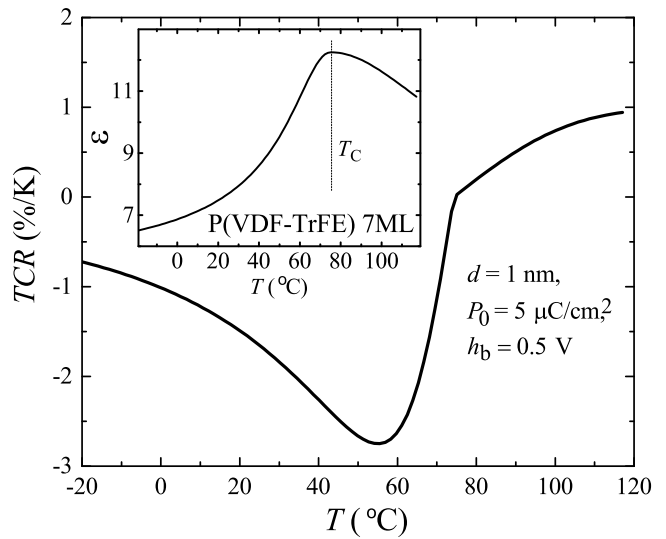}
\caption{(Color online) Temperature coefficient of resistance  (TCR)
as a function of temperature for FTJ with P(VDF-TrFE) barrier
with thickness $d=1$ nm and barrier height $h_\mathrm b=0.5$ V. The inset
shows dependence of the dielectric constant of P(VDF-TrFE) vs.
temperature obtained using
experimental data of Ref.~[\onlinecite{Zlatkin1998}].}\label{Fig:TCRvsTVDF}
\end{figure}

To observe a strong temperature dependence of TJ conductance
due to image forces one needs to use FE with low dielectric constant.
Recently, an organic ferroelectric P(VDF-TrFE) was used as a TJ barrier and
the GER effect was demonstrated in this system. P(VDF-TrFE) has rather small
dielectric constant depending on temperature in the vicinity of the Curie
point $T_\mathrm C\approx 75^o$ C. Figure~\ref{Fig:TCRvsTVDF} shows
the TCR for symmetric FTJ with  P(VDF-TrFE) barrier.
We fit the experimental data of Ref.~[\onlinecite{Zlatkin1998}] on P(VDF-TrFE) dielectric
constant as a function of temperature, $T$ (see inset in Fig.~\ref{Fig:TCRvsTVDF}) with Eq.~(\ref{Eq:DielT}).
The magnitude of TCR reaches 3\%. Note that P(VDF-TrFE) has rather small saturation
polarization ($P_0\approx 5~\mu$C/cm$^2$). Therefore, the main contribution to the
TCR is produced by the image forces and surface charges can
be neglected below $T_\mathrm C$. Above $T_\mathrm C$ the
temperature dependence of TJ conductance appears only due to image forces.

\begin{figure}
\includegraphics[width=1\columnwidth]{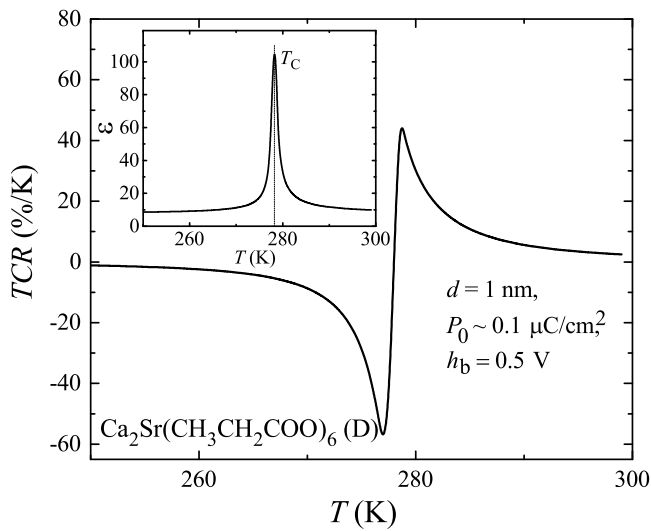}
\caption{(Color online) Temperature resistance coefficient (TCR)
as function of temperature for FTJ with SrCa$_2$Sr(CH$_3$CH$_2$COO)$_6$(D)
barrier with thickness $d=1$ nm and barrier height $h_\mathrm b=0.5$ V.
The inset shows the dependence of the dielectric constant of SrCa$_2$Sr(CH$_3$CH$_2$COO)$_6$(D)
from Ref.~[\onlinecite{Xiong2012}].}\label{Fig:TCRvsTCaSr}
\end{figure}

There are numerous organic FEs with low dielectric constant
($\epsi<100$, see review paper~[\onlinecite{Xiong2012}]). Usually these FEs
have a very small saturation polarization ($P_0\sim 0.1~\mu$C/cm$^2$). Such organic FEs
can be promising candidates for TJ with high TCR. These FEs have the
Curie temperature in a wide range from 50 K (TTF-BA) to a room temperature
(Ca$_2$Pb(CH$_3$CH$_2$COO)$_6$). In some of these FEs the dielectric constant
changes strongly (from 10 to 100) in a very narrow temperature
range ($\Delta T=25$ K) leading to large TCR of order of 50\%/K
(for example Ca$_2$Ba(CH$_3$CH$_2$COO)$_6$, see Fig.~\ref{Fig:TCRvsTCaSr}).
Some of these FEs have a wide peak of dielectric constant
around $T_\mathrm C$ leading to moderate TCR in a wide temperature range.

\section{Conclusion}

We studied influence of image forces on the ER effect and conductance temperature
dependence in FTJs. Image forces inside the FTJ barrier reduce the average
barrier height and strongly influence the TJ conductance. These forces produce the
ER effect at non-zero bias. The effect appears due to dependence of the
FE dielectric constant on the applied bias. For symmetrical FTJ (with
identical metal electrodes) the ER effect due to image forces may exceed
the ER effect due to surface charges at the FE/metal interfaces.
The ER effect due to image forces increases with decreasing
the barrier height and average barrier dielectric constant and
almost independent of the barrier thickness. The importance of
this mechanism (in comparison to the surface charges mechanism) grows with
increasing of barrier height and decreasing of saturation polarization
and barrier thickness. The magnitude of the effect for HfZrO$_2$ FE reaches 50\%.

For strongly asymmetric barrier the contribution of image forces to the
ER effect is small in comparison to the effect of surface charges.
The contribution of the image forces to the ER effect is visible
only for TJ with metallic leads where difference
of Fermi levels does not exceed 50\%.

We studied temperature dependence of the FTJ conductance by taking into
account the image forces. Above the FE Curie point the image forces is
the only mechanism for dependence of the TJ conductance on temperature.
Below $T_\mathrm C$ both the surface charges and the image forces contribute
to the temperature dependence of conductance. Large TCR can be achieved in FTJ
with FE with low dielectric constant in the vicinity of the
FE phase transition. We calculated the TCR for FTJ with P(VDF-TrFE) barriers. The
peak value of TCR is about 3\%/K which is comparable with TCR obtained in
asymmetric FTJ with BTO barrier. According to our analysis the best materials
for observing the strong temperature dependence of TJ conductance are
organic FEs. These materials have a low dielectric constant with strong
relative variation. For example, TJ with SrCa$_2$Sr(CH$_3$CH$_2$COO)$_6$(D)
barrier leads to TCR up to 50\% in the narrow temperature range.

\section{Acknowledgements}
This research was supported by NSF under Cooperative Agreement Award EEC-1160504,
the U.S. Civilian Research and Development Foundation (CRDF Global)
and NSF PREM Award. O.U. was supported by Russian Science Foundation (Grant  16-12-10340).

\bibliography{FTJ}

\end{document}